\documentclass[aps,preprint,showpacs,preprintnumbers,amsmath,amssymb,superscriptaddress]{revtex4}

\usepackage{amsmath}
\usepackage{amssymb}
\usepackage[dvips]{graphicx}

\begin{document}
%\draft

\title{$^{63/65}$Cu- and $^{35/37}$Cl-NMR Studies of Triplet Localization in the Quantum Spin System NH$_4$CuCl$_3$}

\author{H. Inoue$^1$, S. Tani$^1$, S. Hosoya$^1$, K. Inokuchi$^1$, T. Fujiwara$^1$, T. Saito$^1$, T. Suzuki$^{2,1}$, A. Oosawa$^1$, T. Goto$^1$, M. Fujisawa$^3$, H. Tanaka$^3$, T. Sasaki$^4$, S. Awaji$^4$, K. Watanabe$^4$ and N. Kobayashi$^4$}

\affiliation{
$^1$Department of Physics, Sophia University, Kioi-cho, Chiyoda-ku, Tokyo 102-8554, Japan \\
$^2$Advanced Meson Science Laboratory, RIKEN Nishina Center, Hirosawa, Wako-shi, Saitama 351-0198, Japan \\
$^3$Department of Physics, Tokyo Institute of Technology, Oh-okayama, Meguro-ku, Tokyo 152-8551, Japan \\
$^4$Institute for Materials Research, Tohoku University, Katahira, Aoba-ku, Sendai 980-8577, Japan \\
}

\date{\today}

\begin{abstract}
$^{63/65}$Cu- and $^{35/37}$Cl-NMR experiments were performed to investigate triplet localization 
in the $S=1/2$ dimer compound NH$_4$CuCl$_3$, which shows magnetization plateaus at one-quarter and 
three-quarters of the saturation magnetization. 
In $^{63/65}$Cu-NMR experiments, signal from only the singlet Cu site was observed, 
because that from the triplet Cu site was invisible due to the strong spin fluctuation of onsite 3$d$-spins. 
We found that the temperature dependence of the shift of $^{63/65}$Cu-NMR spectra 
at the singlet Cu site deviated from that of macroscopic magnetization below $T=6$ K. 
This deviation is interpreted as the triplet localization in this system.
From the $^{35/37}$Cl-NMR experiments at the 1/4-plateau phase, 
we found the two different temperature dependences of Cl-shift, 
namely the temperature dependence of one deviates below $T=6$ K from that of the macroscopic magnetization 
as observed in the $^{63/65}$Cu-NMR experiments, 
whereas the other corresponds well with that of the macroscopic magnetization 
in the entire experimental temperature region. 
We interpreted these dependences as reflecting the transferred hyperfine field at the Cl site 
located at a singlet site and at a triplet site, respectively.
This result also indicates that the triplets are localized at low temperatures. 
$^{63/65}$Cu-NMR experiments performed at high magnetic fields between the one-quarter 
and three-quarters magnetization plateaus have revealed
that the two differently oriented dimers in the unit cell are equally occupied by triplets,
the fact of which limits the theoretical model on the periodic structure of the localized triplets.
\end{abstract}

\pacs{76.60.-k, 75.10.Jm}

\maketitle

\section{Introduction}

In last two decades, quantum spin systems have been attracting much interests both experimentally and theoretically, 
because these systems exhibit numerous peculiar magnetic features, which cannot be interpreted by conventional classical spin models. 
For instance, a spin gap \cite{Dagotto} 
and a step like magnetization process (magnetization plateau) corresponding to the quantization of the magnetization \cite{Rice} are the macroscopic quantum phenomena. 
When a magnetic field is applied in a spin gap system with the excitation gap $\Delta$, the excited triplet states split and the energy of the one of the triplet states is lowered due to the Zeeman interaction so that the spin gap vanishes at the critical field $H_c = \Delta / g \mu_{\rm B}$. 
Then the ground state becomes magnetic so that the magnetic ordering occurs if there are three-dimensional interactions. 
Such field-induced magnetic ordering has been observed in the spin gap systems KCuCl$_3$ \cite{OosawaK} and TlCuCl$_3$ \cite{OosawaTl}, which are isomorphous of the title compound NH$_4$CuCl$_3$. 
The field-induced magnetic ordering was captured as the Bose-Einstein condensation of magnons by mapping the spin gap system to the system
consisting of boson with magnetic moment, {\it magnon} \cite{Nikuni}. 
Namely the ordering can be interpreted as the superfluid-insulator transition induced by the varying of the chemical potential $\mu$ corresponding to the magnetic field $H$ in the original spin system. 
It is considered that the superfluid-insulator transition occurs when the hopping of magnon $t$ is more dominant than the repulsive interaction of magnons $U$ and this situation seems to be valid for KCuCl$_3$ and TlCuCl$_3$. 
On the other hand, when the repulsive interaction of magnons $U$ is more dominant than the hopping of magnon $t$, the created magnons prefer to distance themselves so that a superlattice of magnons, which can be interpreted as the formation of the Wigner crystal of magnons, may appear at a fractional number of magnons. 
In the superlattice phase, a finite energy is needed to create an additional magnon, i.e. $d n/d \mu = 0$, where $n$ is the total magnon density corresponding to the magnetization $M$ in the original spin system so that the magnetization plateau emerges in the superlattice phase. 
Actually, such superlattice of magnons has been observed in the orthogonal dimer system SrCu$_2$(BO$_3$)$_2$ \cite{KodamaSCBO} at the 1/8-plateau phase \cite{Kageyama,Onizuka}. \par
\begin{figure}[t]
\begin{center}
\includegraphics[width=85mm]{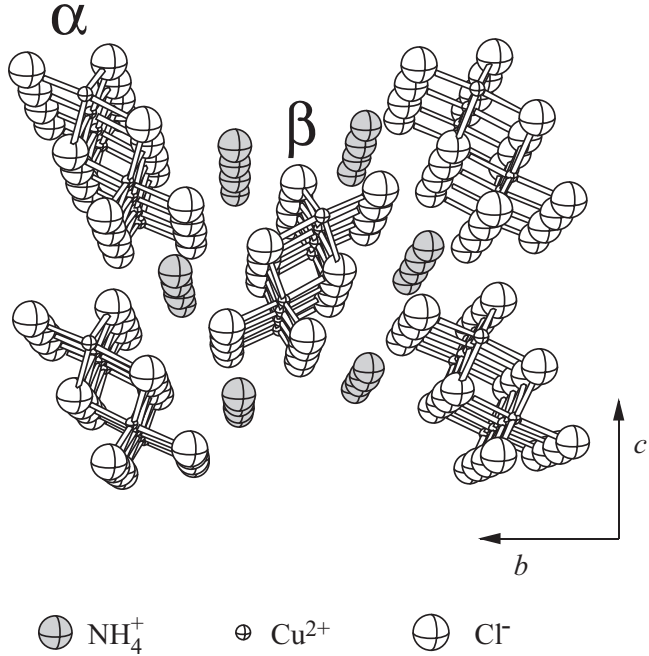}\\
\end{center}
\caption{Crystal structure of NH$_4$CuCl$_3$ viewed along the $a$-axis. Shaded, small open and large open circles denote NH$_4^+$, Cu$^{2+}$ and Cl$^-$ ions, respectively. The differently-oriented Cu$_2$Cl$_6$ dimers at the corners and center of the unit cell are labelled as the $\alpha$ and $\beta$ dimers, respectively. \label{Fig1}\\} 
\end{figure}

The title compound NH$_4$CuCl$_3$ has a monoclinic structure (space group $P2_1/c$) at room temperature \cite{Willett}. The crystal structure is composed of planar dimers of Cu$_2$Cl$_6$. The dimers are stacked on top of one another to form infinite double chains parallel to the crystallographic $a$-axis. These double chains are located at the corners and center of the unit cell in the $b-c$ plane as shown in Fig. \ref{Fig1}. We labelled the differently-oriented Cu$_2$Cl$_6$ dimers at the corners and center of the unit cell as the $\alpha$ and $\beta$ dimers, respectively, in the present paper as shown in Fig. \ref{Fig1}. \par

\begin{figure}[t]
\begin{center}
\includegraphics[width=85mm]{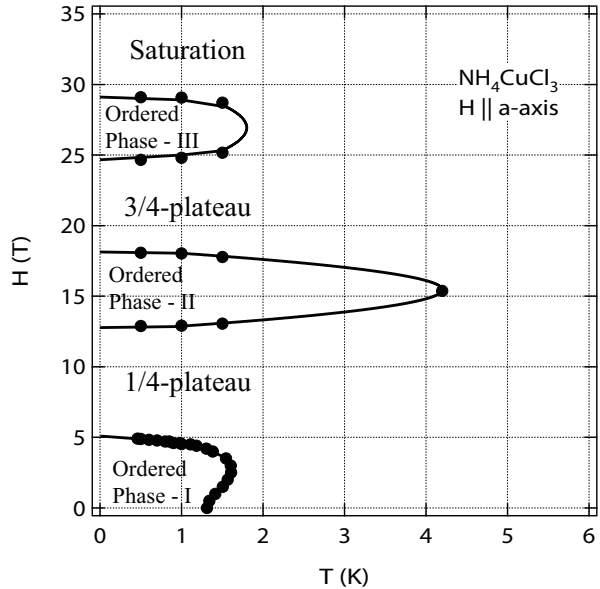}\\
\end{center}
\caption{Phase diagram of NH$_4$CuCl$_3$ for the $H \parallel a$-axis 
determined from the previous specific heat measurements 
\cite{Budhymag,Fujisawa} 
and magnetization measurements \cite{ShiramuraNH4}. 
Solid lines are guides for the eyes. \label{Fig2}}
\end{figure}

The magnetic ground states of the isomorphous compounds KCuCl$_3$ and TlCuCl$_3$ are the spin singlet with excitation gaps \cite{Tanaka, Takatsu, Shiramura, OosawaTl, OosawaK}. From the analyses of the dispersion relations obtained by neutron inelastic scattering experiments, it was found that the origin of the spin gap is the strong antiferromagnetic interaction in the chemical dimer Cu$_2$Cl$_6$, and that the neighboring dimers are coupled by the interdimer interactions along the double chain and in the (1, 0, $-$2) plane \cite{KatoKinela, PSIKinela, PSIKinela2, OosawaTlinela, PSITlinela}. On the other hand, NH$_4$CuCl$_3$ is a gapless antiferromagnet with $T_{\rm N} = 1.3$ K \cite{Budhymag}. NH$_4$CuCl$_3$ presents salient magnetization plateaus in the magnetization process at one-quarter and three-quarters of the saturation magnetization \cite{ShiramuraNH4}, i.e., for $H || a$, these plateaus are observed in 5.0 T $< H <$ 12.8 T, and 17.9 T $< H <$ 24.7 T, respectively, and the magnetization saturates at $H_s$ = 29.1 T. Because the magnetization plateaus are observed irrespective of the field direction, the origin of the plateau can be attributed to quantum effect. Figure \ref{Fig2} shows the phase diagram of NH$_4$CuCl$_3$ determined from the previous specific heat measurements \cite{Budhymag,Fujisawa} and magnetization measurements \cite{ShiramuraNH4}. It can be expected that NH$_4$CuCl$_3$ undergoes three kinds of antiferromagnetic orderings through the one-quarter and three-quarters magnetization plateaus up to the saturation field at low temperature upon increasing the magnetic field. \par

\begin{figure*}[t]
\begin{center}
%%trim=left, bottom, right,up
%%[trim=0.2cm 5.0cm 0.1cm 2.0cm, clip, width=80mm]
\includegraphics[trim=0.2cm 12.0cm 0.1cm 2.0cm,clip, width=120mm]{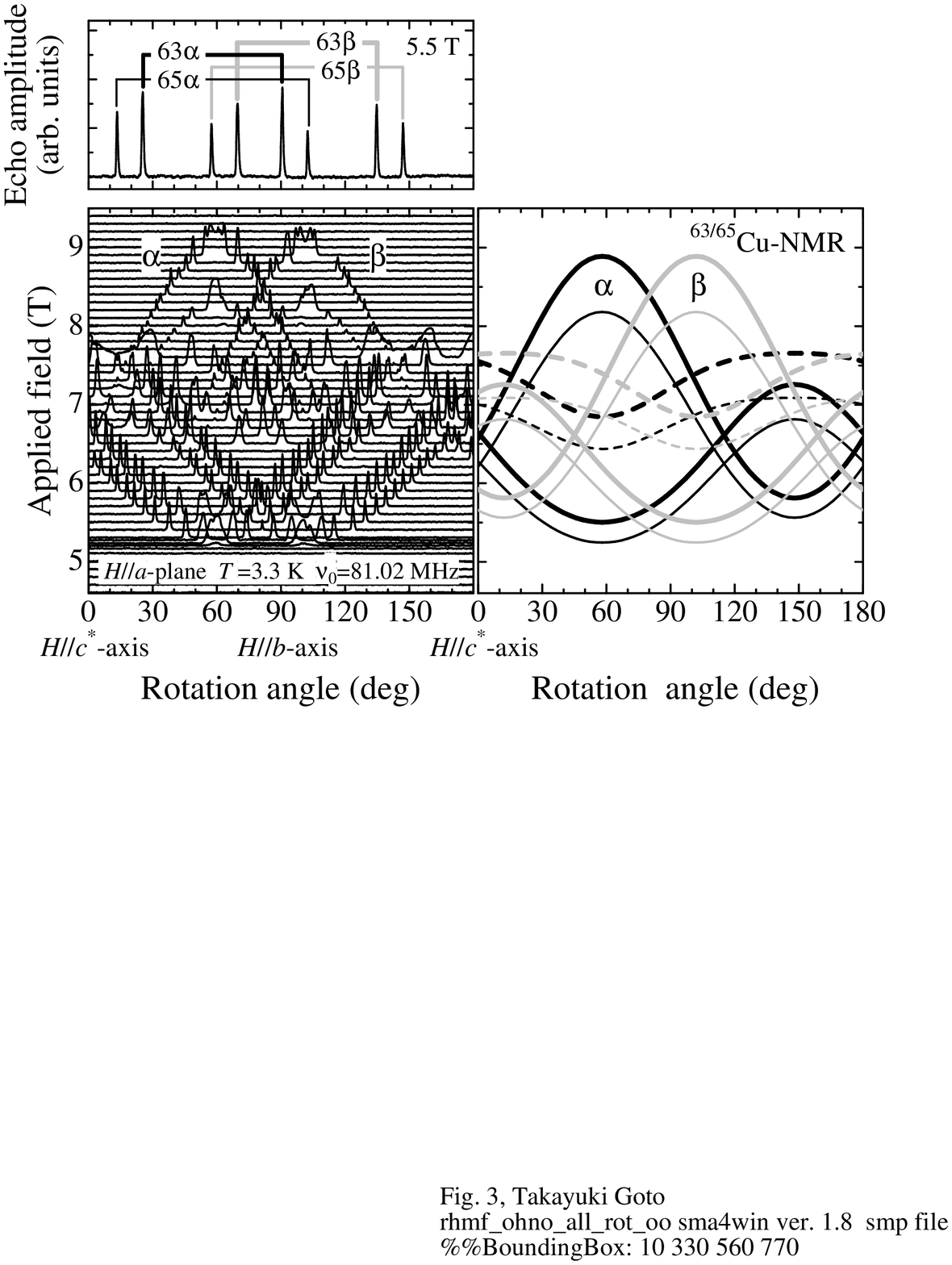}\\
\end{center}
\caption{(Left) Rotational profiles of $^{63/65}$Cu-NMR spectra under the field region of the 1/4-plateau phase 
at $T=3.3$ K in NH$_4$CuCl$_3$. 
The direction of the magnetic field was rotated from $c*$-axis to $b$-axis and then $c^*$-axis, with keeping $H\perp a$-axis.
There are two signal sets, which come from the two differently-oriented dimers $\alpha$ and $\beta$ in a unit cell connected 
with the glide symmetry, as shown in Fig. \ref{Fig1}. 
The upper panel shows the enlarged representative spectrum taken under the constant field 5.5 T, where only
satellite peaks are observed.
(Right) Calculated curves obtained by numerical diagonalization of the nuclear spin Hamiltonian. See text for details. 
\label{Fig3}\\}
\end{figure*}

Recently, Matsumoto \cite{Matsumoto} theoretically suggested that there are three distinct magnetic dimer sublattices 
with different exchange interactions in NH$_4$CuCl$_3$, 
in contrast to the isostructural KCuCl$_3$ and TlCuCl$_3$, and reproduced well the magnetization plateaus 
and field dependence of the magnetic resonance frequencies observed in ESR measurements \cite{BudhyESR, Nojiri} by his model. 
Successive structural phase transitions which may indicate 
the creation of such distinct magnetic dimers have been observed at low temperatures by means of various kind of experiments \cite{Heyns, Schmidt, Ruegg, Kodama, Shimaoka} while the isostructural KCuCl$_3$ and TlCuCl$_3$ undergo no structural phase transitions at low temperature. 
Although some experiments proposed that the crystal structure becomes triclinic space group $P1$ or $P{\bar 1}$ which enables to form 
multiple distinct magnetic dimers by enlarging the unit cell, 
the crystal structure including the detailed atomic positions at low temperature has not been fully determined yet. 
The magnetic ordering at zero field was observed and the magnetic structure at the 1/4-plateau state was proposed by the neutron elastic scattering experiment \cite{Ruegg}. In terms of the NMR experiments, $^{15}$N-NMR \cite{Shimaoka} and $^{14}$N-NMR \cite{Kodama} experiments have been reported so far in NH$_4$CuCl$_3$, in which the N ions exist in the NH$_4$ molecules between the double chains formed by planar dimers of Cu$_2$Cl$_6$ as shown in Fig. \ref{Fig1}. 
In $^{15}$N-NMR experiment, the periodic structure of the triplets at the 1/4-plateau phase was proposed from the calculation of the internal fields at N sites by using a point dipole approximation. Also, in $^{14}$N-NMR experiment, the periodic structure at the 1/4-plateau phase was proposed from the discussion of the crystal symmetry at low temperature. However, there remain controversies in the magnetic structure at zero field together with the periodic structure of the triplets at the plateau phases, as well as the crystal structure. \par
As mentioned above, the periodic structure of the triplets has not been determined yet. Moreover, the triplet localization at low temperature has not been observed directly yet. In order to observe the triplet localization, it is important to perform NMR experiments of Cu and Cl ions, which are closer to the magnetic moments at Cu sites than N ions, as shown in Fig. \ref{Fig1}, in NH$_4$CuCl$_3$. With this motivation, we performed the $^{63/65}$Cu- and $^{35/37}$Cl-NMR experiments at the 1/4-plateau phase in the quantum spin system NH$_4$CuCl$_3$. As the result, we actually observed the triplet localization, indicating the existence of the inequivalent magnetic dimer sites, at low temperature in NH$_4$CuCl$_3$ for the first time, as presented in the following sections. Also, we carried out the $^{63/65}$Cu-NMR experiments at high magnetic fields between the 1/4-plateau and the 3/4-plateau phases in order to investigate the periodic structure of the localized triplets in NH$_4$CuCl$_3$ and will discuss the observed results with the periodic structure of the localized triplets proposed in the previous neutron \cite{Ruegg} and N-NMR \cite{Shimaoka,Kodama} experiments from the viewpoint of the temperature and magnetic field dependence of the amplitudes of the observed $^{63/65}$Cu-NMR spectra in the following sections. \par

\section{Experimental Details}

Single crystals of NH$_4$CuCl$_3$ were prepared by a slow evaporation method \cite{ShiramuraNH4}. $^{63/65}$Cu- and $^{35/37}$Cl-NMR experiments were performed at temperatures down to $T=2$ K  in magnetic fields up to $H=18$ T using the 12 T cryogen-free-superconducting magnet installed at Sophia University and the 20 T superconducting magnet in the High Field Laboratory for Superconducting Materials, Institute for Materials Research, Tohoku University. NMR rotational spectra were obtained by recording the spin-echo amplitude while rotating the crystal very slowly in the magnetic field. For the estimation of hyperfine coupling constants, the macroscopic magnetization was measured using a SQUID magnetometer with a 7 T superconducting magnet (Quantum Design, MPMS-XL) and a vibrating sample magnetometer with a 14 T superconducting magnet (Oxford, MagLabVSM) in the Center for Low Temperature Science, Tohoku University.  

\section{Results and Discussion}

\subsection{$^{63/65}$Cu-NMR Study at Low Magnetic Fields}

The rotational profiles of $^{63/65}$Cu-NMR spectra under the field region of the 1/4-plateau phase at $T=3.3$ K 
in NH$_4$CuCl$_3$ are shown in Fig. \ref{Fig3}. The direction of the magnetic field is rotated within $a$-plane. 
The curves in the right panel of Fig. \ref{Fig3} were calculated by numerical diagonalization 
of the nuclear spin Hamiltonian of $^{63/65}$Cu ($I=3/2$)
\begin{eqnarray}
\label{eq1}
{\cal H} &=& - \hbar \gamma {\pmb I} \cdot {\pmb H} - \hbar \gamma {\pmb I} \cdot K {\pmb H} \nonumber \\
    & & {} + \frac{h \nu_Q}{6} \left[ \left\{ 3 I_z^2 - I \left( I+1 \right) \right\} + \eta \left(I_x^2 - I_y^2 \right) \right]
\end{eqnarray}
with the parameters of Knight shift $K$$\simeq$0 \%
quadrupole interaction parameter $^{63}\nu_Q$=39.2 MHz, and asymmetry parameter $\eta$=0, 
and with the principal axis of the field gradient tensor directed perpendicular to the basal plane of the Cu$_2$Cl$_6$ octahedron. 
Black and gray curves correspond to the two differently-oriented dimers $\alpha$ and $\beta$ in a unit cell as shown in Fig, \ref{Fig1}.
thick and thin lines correspond to the two isotopes of $^{63}$Cu and $^{65}$Cu, and solid and dashed lines correspond to the two satellite transitions and the center transition. The agreement between the observed data and the calculation with nearly zero shift indicates that the observed signal comes from the singlet site. Reflecting the fact that there are two differently-oriented dimers $\alpha$ and $\beta$ connected with the glide symmetry in a unit cell, which is expected from the $P2_1/c$ crystal symmetry, as shown in Fig. \ref{Fig1}, there are two sets of signals separated by approximately 40$^\circ$, which corresponds to the angle between the normals of the basal planes of $\alpha$ and $\beta$ projected onto the $a$-plane. These two sets of signals in Fig. \ref{Fig3} have nearly the same amplitude. This result indicates that there are singlet dimers in both the $\alpha$ and $\beta$ sites in the 1/4-plateau region. \par
As mentioned in the Introduction, because the crystal symmetry is lowered at low temperature in this system, 
splitting of Cu-NMR spectra to reflect the lowered crystal structure is expected. However, 
the observed Cu-NMR spectra was reproduced by the calculation on the assumption that the crystal structure 
belongs to the $P2_1/c$ space group, as shown in Fig. \ref{Fig3}. 
This indicates that the symmetry of the local crystal field at Cu sites preserves the $P2_1/c$ symmetry even at low temperature, 
and that the lowering of crystal symmetry is associated with the variation of the symmetry between the NH$_4$ molecules, 
such as the long range order of the orientation of NH$_4$ molecules. 
In this $^{63/65}$Cu-NMR experiment, the signal from the triplet site was not observed, because of the strong spin fluctuation of onsite 3$d$-spins. \cite{Hosoya} \par
%%trim=left, bottom, right,up
%%[trim=0.2cm 12.0cm 0.1cm 2.0cm, clip, width=80mm]
\begin{figure}[t]
\begin{center}
\includegraphics[trim=0cm 0cm 0cm 0cm, clip, width=90mm]{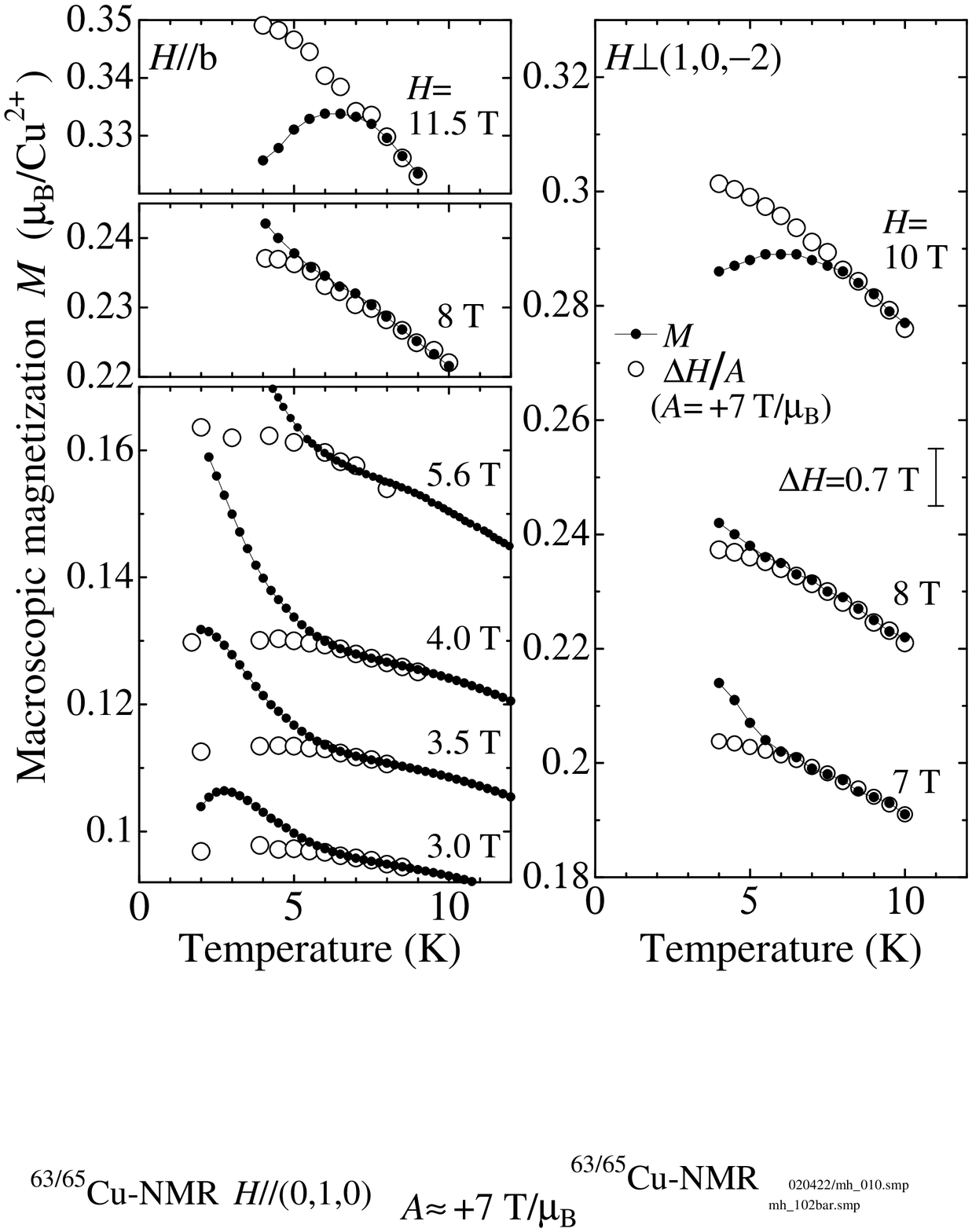}\\
\end{center}
\caption{Representative temperature dependences of the relative shift $\Delta H$ of $^{63/65}$Cu-NMR (open circles) and the macroscopic magnetization $M$ (closed circles) at various magnetic fields for the $H \parallel b$-axis and $H \perp (1, 0, -2)$ plane. The former are scaled as $\Delta H / A$, where $A$ is the hyperfine coupling constant with $+7$ T$/\mu_{\rm B}$. \\
\label{Fig4}}
\end{figure}
\begin{figure}[t]
\begin{center}
\includegraphics[trim=0cm 8cm 0cm 3cm, clip, width=100mm]{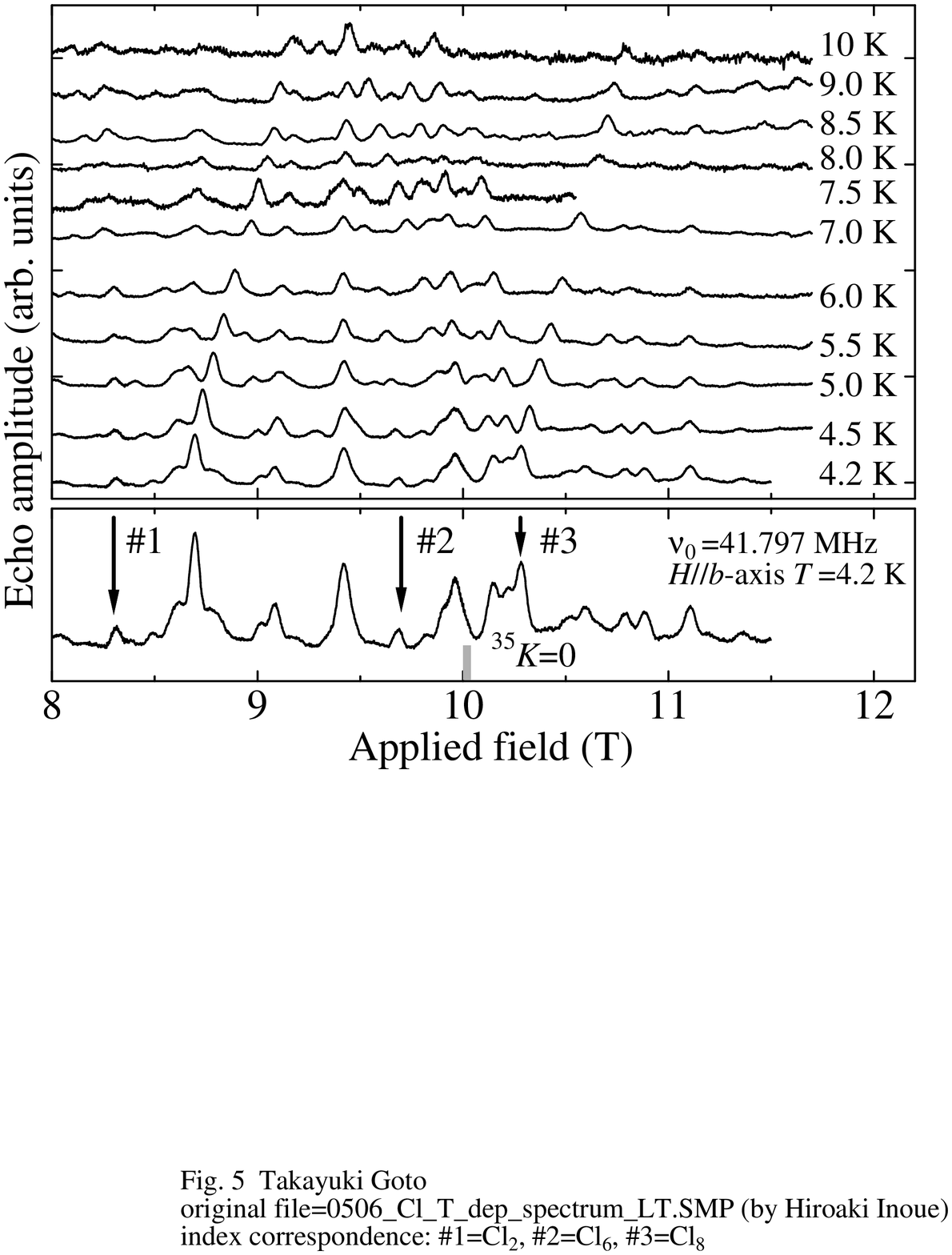}\\
\end{center}
\caption{(Upper panel) Magnetic field-swept spectra of $^{35/37}$Cl-NMR 
at various temperatures for the $H \parallel b$-axis in NH$_4$CuCl$_3$. 
(Lower panel) Enlarged figure of magnetic field-swept spectra at $T=4.2$ K 
for the $H \parallel b$-axis.
The three peaks numbered as \#1, 2 and 3 are those from $^{35}$Cl-nuclei, 
and the temperature dependence of their shift is shown in Fig. \ref{Fig6}. 
The gray bar denotes the zero-shift position of $^{35}$Cl for the present 
frequency $\nu_0=$41.797 MHz. \\
\label{Fig5}
}
\end{figure}
Representative temperature dependences of the relative shift $\Delta H$ of $^{63/65}$Cu-NMR and the macroscopic magnetization $M$ at various magnetic fields for the $H \parallel b$-axis and the $H \perp (1, 0, -2)$ plane are shown in Fig. \ref{Fig4}. We can see a remarkable deviation in the two quantities below $T=6$ K. 
The hyperfine coupling constant $A$ determined in the temperature range above $T=6$ K, where the two quantities are proportional, is about $A\simeq$+7 T$/\mu_{\rm B}$, which is positive and very small. 
Divalent Cu atoms generally have a large and negative hyperfine coupling of a few tens of tesla, that is $A<0$, because of the core-polarization effect. 
Therefore, the observed positive value of $A$ means that the local magnetization probed by Cu-NMR as the shift and the macroscopic magnetization $M$ must show opposite temperature dependences. 
Note that this anomalous behavior is not merely due to the dipole anisotropy, but originates in the electron spin state, 
because $A$ takes the same value irrespective of the field directions parallel or perpendicular to the Cu$_2$Cl$_6$ basal plane, 
as shown in the left and right panels of Fig. \ref{Fig4}. 

%%trim=left, bottom, right,up
%%[trim=0.2cm 12.0cm 0.1cm 2.0cm, clip, width=80mm]
\begin{figure*}[t]
\begin{center}
\includegraphics[trim=0cm 0cm 0cm 0cm, clip, width=100mm]{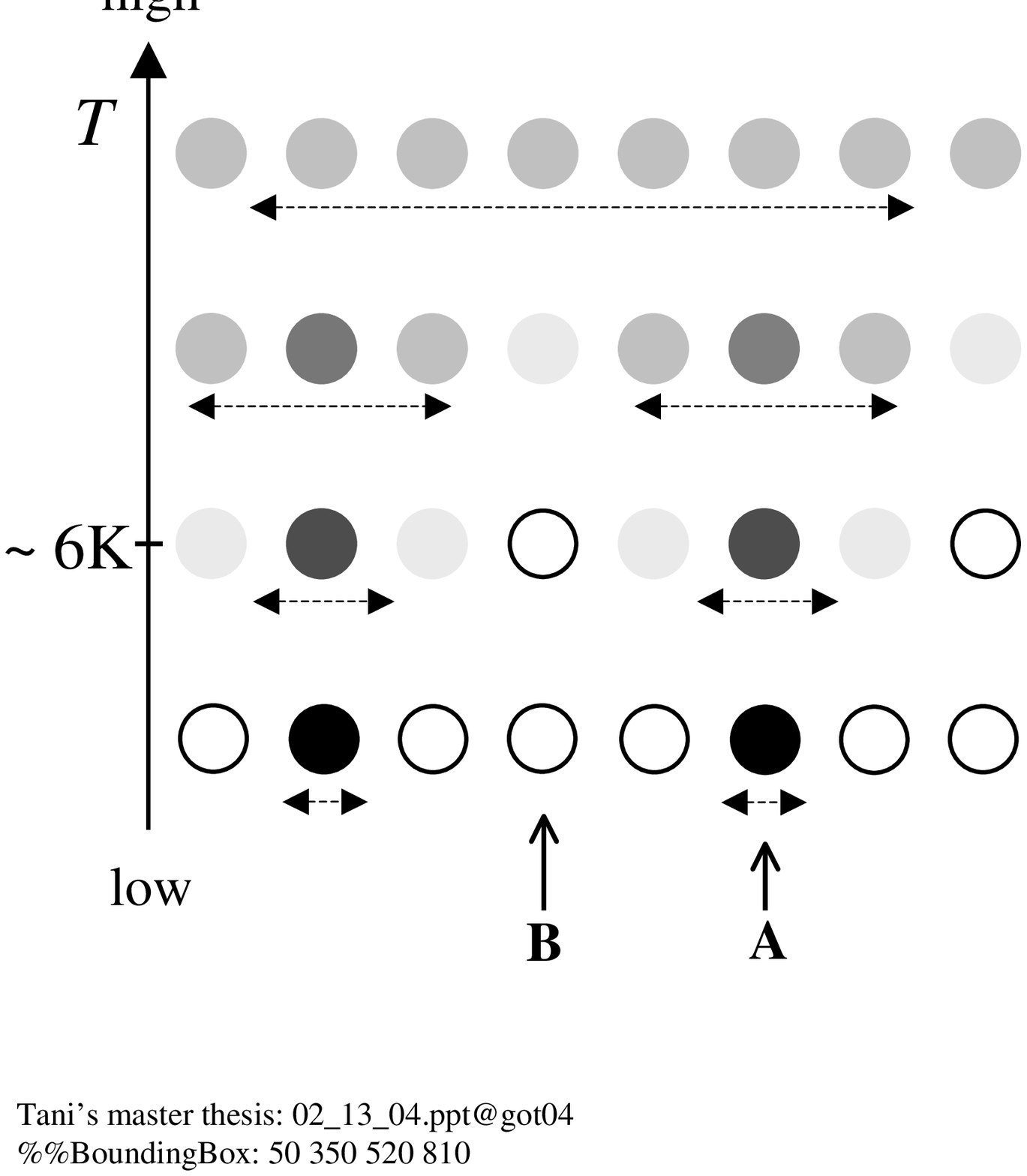}\\
\end{center}
\caption{Schematic picture of the triplet localization. 
A Circle in each row corresponds to the dimer site in NH$_4$CuCl$_3$ at each temperature.
Upper and lower rows corresponds to the higher and lower temperatures respectively.
The shading of each circle is proportional to the local magnetization of the dimer.  
Horizontal axes show the localization length at each temperature.
The two vertical axes A and B
at the bottom show representative positions where the triplet
stays or does not at low temperatures.
\\
\label{Fig6}}
\end{figure*}

To explain this phenomenon, we propose a model in which the field-induced triplets are spatially localized and contribute only 
to the macroscopic magnetization and not to the local magnetization of neighboring singlet sites. 
Its schematic drawing is shown in Fig. \ref{Fig6}.
In the high temperature limit, those triplets are mobile and the system is homogeneous. 
As the temperature is lowered below a few tens of kelvins, 
they start to become localized and tend to stop at some specific spatial positions, indicated as {\bf A} 
and avoid to stop at some other positions, indicated as {\bf B}
in Fig. \ref{Fig6}.
Since the local magnetic field at the singlet site {\bf B} is produced only by moving triplets, 
it always decreases with decreasing triplet localization length, and hence with decreasing temperature. 
This agrees with the positive and small $A$ in Fig. \ref{Fig4} above $T=6$ K. Below $T=6$ K, 
the localization length must be shortened to be less than the mean dimer-dimer separation, 
so that the local magnetization completely saturates. 

This localization model is consistent with the neutron inelastic scattering experiments 
reporting a very small dispersion of triplet excitations in NH$_4$CuCl$_3$ \cite{OosawaNH4inela} and 
indicates the experimental evidence of the existence of inequivalent magnetic dimer sites at low temperature, 
because it can be considered that the triplets scarcely hop to inequivalent magnetic dimer sites 
with different exchange interactions. 
This model predicts that the temperature dependence of the local magnetization of the triplet site, 
indicated as {\bf A} in Fig. \ref{Fig6}, must agree with the macroscopic $M$, which is to be discussed
in the following subsection.

\subsection{$^{35/37}$Cl-NMR Study}

%%trim=left, bottom, right,up
%%[trim=0.2cm 12.0cm 0.1cm 4.0cm, clip, width=100mm]
\begin{figure*}[t]
\begin{center}
\includegraphics[trim=0cm 11cm 1cm 1cm, clip, width=130mm]{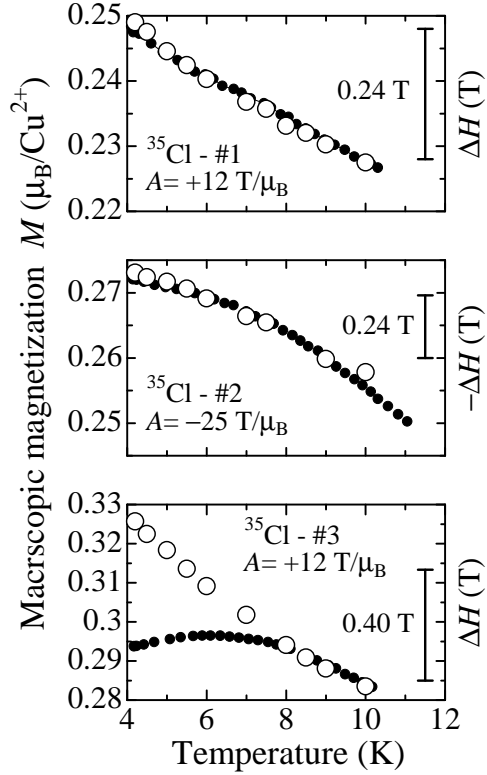}\\
\end{center}
\caption{
Representative temperature dependences of the relative shift $\Delta H$ of 
$^{35/37}$Cl-NMR (open circles) and the macroscopic magnetization $M$ (closed circles) for peaks \#1, 2 and 3 
defined in Fig. \ref{Fig5}. 
The former are scaled as $\Delta H / A$, where $A$ is the hyperfine coupling constant. 
Note that $\Delta H$ of Cl-\#2 with negative $A$ actually decreases with decreasing temperature. 
\label{Fig7}
\\}
\end{figure*}

%%trim=left, bottom, right,up
%%[trim=0.2cm 13.0cm 0.1cm 2.0cm, clip, width=80mm]
\begin{figure*}[t]
\begin{center}
\includegraphics[trim=0cm 14cm 0cm 3cm, clip, width=140mm]{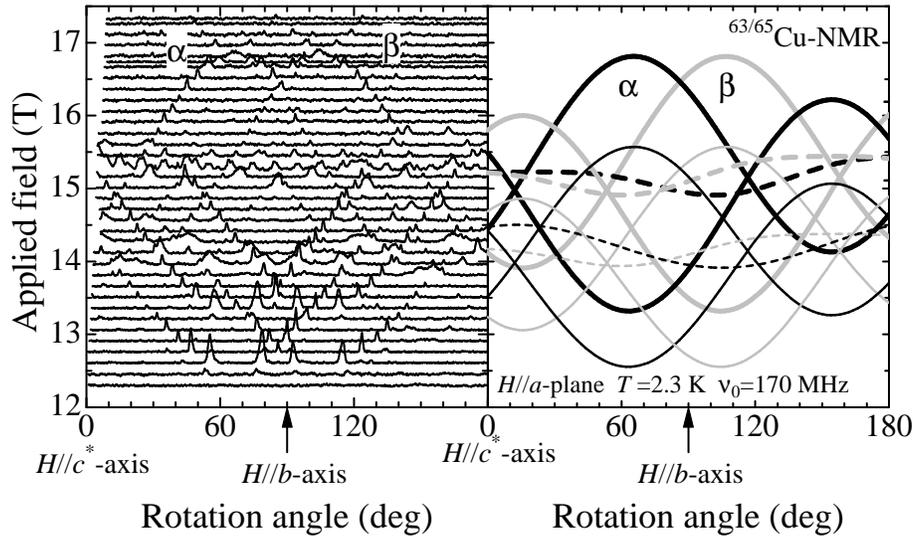}\\
\end{center}
\caption{(Left) Rotational profiles of $^{63/65}$Cu-NMR spectra under the field region between the one-quarter and the three-quarters magnetization plateaus for the $H \parallel a$-plane at $T=2.3$ K in NH$_4$CuCl$_3$. There are two signal sets, which come from the two differently-oriented dimers $\alpha$ and $\beta$ in a unit cell connected with glide symmetry, as shown in Fig. \ref{Fig1}. (Right) Calculated curves obtained by numerical diagonalization of the nuclear spin Hamiltonian. See text for details. 
\\
\label{Fig8}}
\end{figure*}

The magnetic field-swept spectra of $^{35/37}$Cl-NMR at various temperatures for 
the $H \parallel b$-axis in NH$_4$CuCl$_3$ are shown in Fig. \ref{Fig5}. 
There are many peaks of central ($- 1/2 \leftrightarrow + 1/2 $) and 
satellite transitions ($ \pm 3/2 \leftrightarrow \pm 1/2 $) from $^{35/37}$Cl ($I=3/2$) sites. 
Typical temperature dependences of the relative shift $\Delta H$ of $^{35}$Cl-NMR and the macroscopic 
magnetization $M$ are shown in Fig. \ref{Fig7}. 
Here, the temperature dependence of the relative shift $\Delta H$ 
of peak \#3 deviates below $T=6$ K from that of 
the macroscopic magnetization $M$ as observed 
in the $^{63/65}$Cu-NMR experiments described above, 
while the temperature dependences of the relative shifts $\Delta H$ 
of peaks \#1 and 2 correspond well to that of the macroscopic 
magnetization $M$ in the entire experimental temperature region
as $\Delta H=A M$, 
where the hyperfine coupling constant $A$
is obtained as $+12$ and $-25$ ${\rm T}/\mu_{\rm B}$, respectively. 

As mentioned in the previous section on $^{63/65}$Cu-NMR experiments, 
field-induced triplet dimers localize at low temperatures in NH$_{4}$CuCl$_{3}$. 
Based on this consideration, we can interpret the observed results of $^{35/37}$Cl-NMR experiments. 
The appearance of the two kinds of the temperature dependence in the Cl-shift indicates 
that there are two magnetically-inequivalent dimers that are localized. 
As for the Cl sites located at singlet dimer, 
since they are scarcely affected by the neighboring localized triplets,
the temperature dependence of the shift of such Cl site 
should be independent of the macroscopic magnetization, 
which is dominated by the triplets.
This argument is the same as that of the Cu-shift of the singlet sites
given in the previous section. 
On the other hand, 
the local magnetization of Cl sites located at triplet dimer is directly affected
by the transferred hyperfine interaction, 
so the temperature dependence of the shift of such Cl site agrees well with the macroscopic magnetization. 
Hence, we can expect that the observation of the two kinds of 
temperature dependence in the Cl-shift will provide microscopic 
evidence of triplet localization below $T=6$ K in NH$_4$CuCl$_3$. 
The wide range of the hyperfine coupling constants for Cl at each of the two sites is simply 
because the hyperfine interaction at the Cl site is anisotropic.  

%%We show this model as a schematic drawing in Fig. \ref{Fig7}.
%%At high temperatures well above 6 K, triplets move freely in crystal, 
%%leaving an average local magnetization at each dimer site.
%%When the temperature is lowered, 
%%there appear some specific sites where triplets tend to stay at (A) or contrarily avoid
%%to stay at (B).
%%Thus, the local magnetization of the site A increases and the site B decreases as the
%%localization proceeds at low temperatures. 
%%This is why we observed the two different temperature dependences in the local magnetization.
%%The temperature dependence of the macroscopic magnetization $M$
%%is expected to agree with that of A,
%%because $M$ is dominated by those dimer sites where triplets stay at.
The origin of this triplet localization 
is not still clear, but possibly, inhomogenization of the system due to 
the rotation freezing of the non-centrosymmetric
NH$_4$ molecule may be related.\cite{Kodama}
In order to confirm the validity of our localization model, we propose an experiment
on the detection of the slow dynamics in localizing triplets.
This can be probed by muon spin relaxation experiment in the longitudinal field $H_{\rm LF}$
(LF-$\mu$SR).\cite{Goto}
The muon spin relaxation rate $\lambda$ under $H_{\rm LF}$ corresponds to the Fourier
component of $\omega=\gamma H_{LF}$, where $\gamma$ is the gyromagnetic
ration of muon spin, in the spectrum of magnetic field fluctuation produced by triplets.
As the localization proceeds, motion of the triplet is expected to slow down,
and hence the $\lambda$ under specific $H_{\rm LF}$ is expected to be enhanced.
This experiment is now in preparation.

%%The value of $A$ such as 25 ${\rm T}/\mu_{\rm B}$ seems to be very large,
%%considering the fact that the Cl site is a ligand site.
%%At this stage, however, we cannot determine its origin. 

\subsection{$^{63/65}$Cu-NMR Study at High Magnetic Fields}

Rotational profiles of $^{63/65}$Cu-NMR spectra under a field region between the one-quarter 
and the three-quarters magnetization plateaus for the $H \parallel a$-plane 
at $T=2.3$ K in NH$_4$CuCl$_3$ are shown in Fig. \ref{Fig8}. 
We can see that the two sets of signals belonging to two differently-oriented dimers $\alpha$ and $\beta$ 
have nearly the same amplitude. 
The curves in the right panel of Fig. \ref{Fig8} were calculated by numerical diagonalization of 
the nuclear spin Hamiltonian (eq. \ref{eq1}) of $^{63/65}$Cu 
with the parameters of quadrupole interaction parameter $^{63}\nu_Q$=39.2 MHz, 
asymmetry parameter $\eta$=0, 
and Knight shift $K$$\simeq$0 \% and with the principal axis of the field gradient tensor 
directed perpendicular to the basal plane of the Cu$_2$Cl$_6$ octahedron. 
Black and gray curves correspond to the two differently-oriented dimers $\alpha$ and $\beta$ in a unit cell, 
thick and thin lines correspond to the two isotopes of $^{63}$Cu and $^{65}$Cu, 
and solid and dashed lines correspond to the two satellite transitions and the center transition. 
The good agreement between the observed data and the calculation with nearly zero shift indicates that the observed signal came from the singlet site, the same as in the $^{63/65}$Cu-NMR experiments at low magnetic field presented in the previous section. \par
\begin{figure*}[t]
\begin{center}
\includegraphics[trim=0cm 16cm 5cm 1cm, clip, width=100mm]{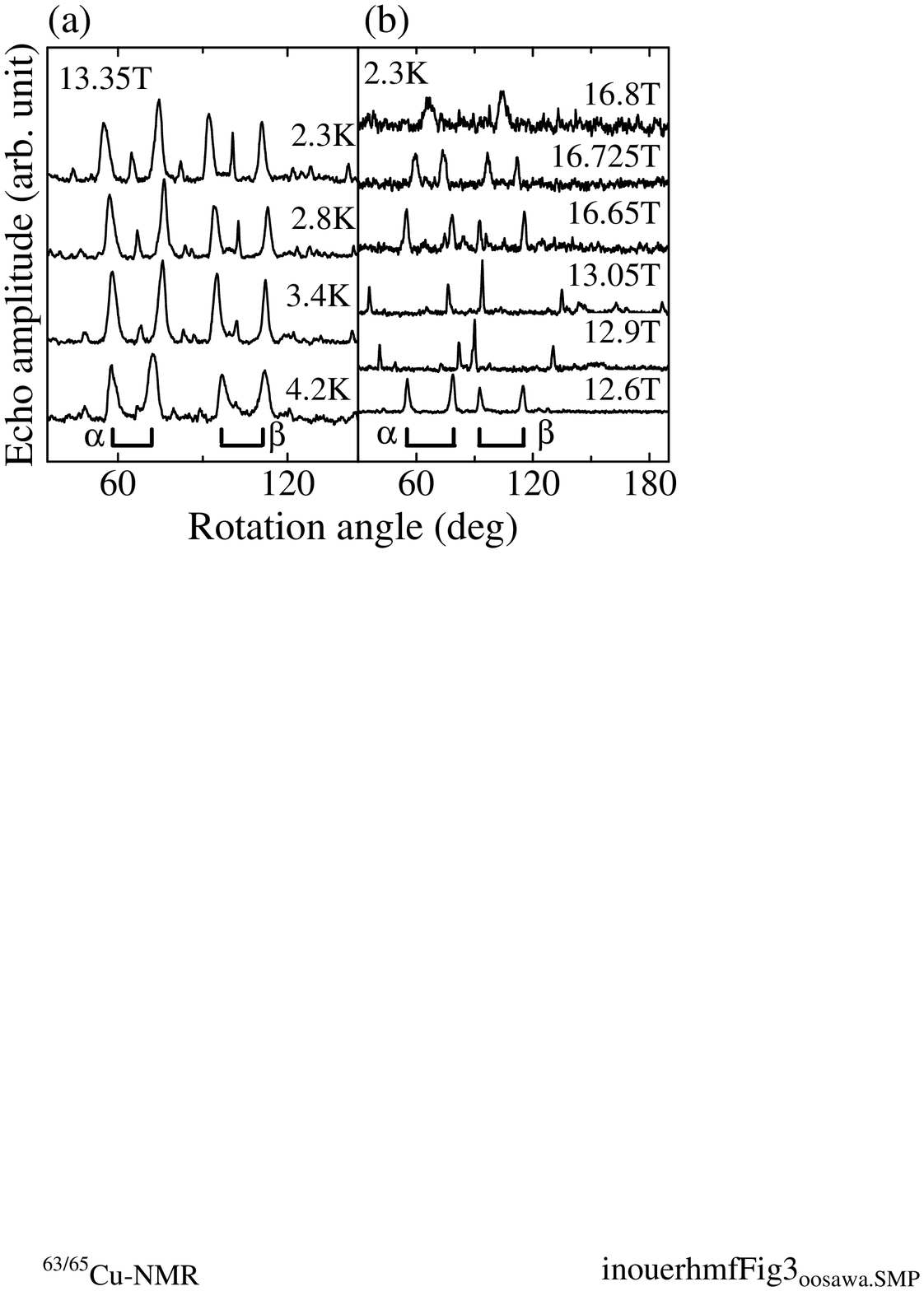}\\
%%%
%%% (a) \hspace{3cm} (b)
%%%
\end{center}
\caption{Representative rotational profiles of $^{63/65}$Cu-NMR spectra of NH$_4$CuCl$_3$ for (a) various temperatures at $H=13.35$ T and (b) various magnetic fields at $T=2.3$ K. \\
\label{Fig9}}
\end{figure*}

Representative rotational profiles of $^{63/65}$Cu-NMR spectra of NH$_4$CuCl$_3$ for various temperatures at $H=13.35$ T and for various magnetic fields at $T=2.3$ K are shown in Figs. \ref{Fig9} (a) and (b), respectively. 
The amplitudes of $^{63/65}$Cu-NMR spectra of $\alpha$ and $\beta$ dimers are almost
the same and show no temperature or magnetic field dependence, 
indicating that there are the same number of singlet sites in both $\alpha$ and $\beta$ dimers 
between the 1/4-plateau and 3/4-plateau phases. 
We observed that the extra peaks indicative of antiferromagnetic ordering appeared at low temperature, as shown in Fig. \ref{Fig9}. 
Because we did not obtain sufficient data to determine the magnetic structure of the ordered phase in the present experiments, 
we do not discuss the observed extra peaks in this paper. \par

%%trim=left, bottom, right,up
%%[trim=0.2cm 12.0cm 0.1cm 2.0cm, clip, width=80mm]

In the models proposed from the neutron experiments by R\"{u}egg {\it et al.} \cite{Ruegg}, a doubly elongated unit cell in the direction of the $b$-axis contains four dimers, two of which are equivalent, and hence three inequivalent dimers with different saturation fields. These three, with an abundance ratio of $1:2:1$, are $\alpha$, $\beta$ and $\alpha$ dimers (or $\beta$, $\alpha$ and $\beta$ dimers), respectively. If we follow their models, between the 1/4-plateau and 3/4-plateau phases, then one quarter of the dimer sites would be completely filled by triplets and half of them would be partially filled by triplets; only the remaining quarter of the sites would retain singlets, which means that they should be visible by NMR. Our observations completely contradict their models which indicate that only one of the two sites of dimers $\alpha$ or $\beta$ remains a singlet in this field region. On the other hand, the periodic structure at the 1/4-plateau phase proposed to explain the $^{15}$N-NMR experiment \cite{Shimaoka}, in which the triplets are arranged per four dimers along the $a$-axis and also satisfy translational symmetry along both $b$- and $c$-axes, is consistent with our observation because in their model there are singlets in both the $\alpha$ and $\beta$ dimers. Moreover, the periodic structure at the 1/4-plateau phase proposed to explain the $^{14}$N-NMR experiment \cite{Kodama}, in which the triplets are formed over different double chains at the corners and center of the unit cell, is consistent with our observation because the triplets in their model are formed by one of the Cu spins in the $\alpha$ dimer and that in the $\beta$ dimer, so the number of Cu sites forming the remaining singlets is the same in both $\alpha$ and $\beta$ dimers. It is important to perform a detailed investigation of spin correlation functions by means of neutron scattering experiments at high magnetic fields in order to investigate the structure of triplets, namely which Cu spins form the triplet, as well as the periodic structure of triplets at plateau phases. 

\section{Conclusion}

We presented the results of $^{63/65}$Cu- and $^{35/37}$Cl-NMR experiments to investigate triplet localization in the quantum spin system NH$_4$CuCl$_3$. In the $^{63/65}$Cu-NMR experiments for the 1/4-plateau phase, we found that the observed rotational profiles could be well reproduced by using the nuclear spin Hamiltonian (eq. {\ref{eq1}) of $^{63/65}$Cu with the parameters of quadrupole interaction parameter $^{63}\nu_Q$=39.2 MHz, asymmetry parameter $\eta$=0, and Knight shift $K$$\simeq$0 \% and with the principal axis of the field gradient tensor directed perpendicular to the basal plane of the Cu$_2$Cl$_6$ octahedron, indicating that the singlet Cu sites are visible only in the $^{63/65}$Cu-NMR experiments, as shown in Fig. \ref{Fig3}. Moreover, we found that the temperature dependence of the relative shift $\Delta H$ of $^{63/65}$Cu-NMR at the singlet Cu site deviated from that of the macroscopic magnetization $M$ below $T=6$ K, as shown in Fig. \ref{Fig4}. This indicates that the triplet localization actually occurs due to the existence of the inequivalent magnetic dimer sites at low temperature. From the $^{35/37}$Cl-NMR experiments for the 1/4-plateau phase, we found two kinds of the temperature dependence of the Cl-shift: in one, the temperature dependence below $T=6$ K deviates from that of the macroscopic magnetization $M$ as observed in the $^{63/65}$Cu-NMR experiments at the 1/4-plateau phase, whereas in the other, the temperature dependence corresponds well to that of the macroscopic magnetization $M$ in the entire experimental temperature region, as shown in Fig. \ref{Fig7}. 
We interpret these two dependences as reflecting the fluctuating field at the Cl site at a singlet site and that at a triplet site, namely this result also indicates that the triplets are localized at low temperature. To investigate the periodic structure of the localized triplets in NH$_4$CuCl$_3$, we also performed $^{63/65}$Cu-NMR experiments at high magnetic fields between the one-quarter and the three-quarters magnetization plateaus and found that the amplitudes of $^{63/65}$Cu-NMR spectra of $\alpha$ and $\beta$ dimers were almost the same and showed no temperature or magnetic field dependence, indicating that there are the same number of singlet sites in both $\alpha$ and $\beta$ dimers between the 1/4-plateau and 3/4-plateau phases, as shown in Fig. \ref{Fig9}. These results do not support the periodic structure of localized triplets proposed in the previous neutron experiment, while are consistent with that proposed in the previous N-NMR experiments.    

\section*{Acknowledgements}

We acknowledge M. Matsumoto, M. Takigawa and Ch. R\"{u}egg for useful discussions and T. Nojima for the help of the macroscopic magnetization measurements. This work was supported by Grants-in-Aid for Scientific Research on Priority Areas "High Field Spin Science in 100 T" from the Ministry of Education, Science, Sports and Culture of Japan, the Saneyoshi Scholarship Foundation and the Kurata Memorial Hitachi Science and Technology Foundation. A part of this study was performed at the High Field Laboratory for Superconducting Materials, Institute for Materials Research, Tohoku University. \par

\end{document}